# Influence of the growth temperature and annealing on the optical properties of $\{CdO/ZnO\}_{30}$ superlattices

E. Przeździecka, A. Lysak, A. Adhikari, M. Stachowicz, A. Wierzbicka, R. Jakiela, K. Zeinab, P. Sybilski, A. Kozanecki

Institute of Physics, Polish Academy of Sciences, Al. Lotników 32/46, Warsaw, Poland

Abstract

Optical properties of the short period {CdO/ZnO} superlattices grown by plasma assisted MBE were analyzed. The superlattice (SLs) structures were successfully obtained at different growth temperatures from 360 to 550 °C. Interestingly, the growth temperature of the SLs influences quality of multilayers and also optical properties of these structures. After annealing at 900°C by rapid thermal method various defect luminescence located at different energetic positions , were detected, and intensity of luminescence strongly depends on applied growth temperature.

Introduction

Zinc oxide (ZnO), a wide direct band gap 3.37 eV II–VI semiconductor, has been perceived as a promising candidate for many applications in various fields. Particularly, it can be used in transparent optoelectronic devices such as ultraviolet (UV) solid-state emitters and detectors, etc[1–4]. The importance of ZnO has been enhanced by the ongoing band gap engineering research toward fabrication of efficient ZnO-based emitters and detectors with a wide dedicated spectral range[5,6]. The band gap of ZnO can be engineered by alloying of ZnO with cadmium oxide (CdO) or magnesium oxide (MgO). CdO has a direct bandgap of 2.3 eV at Γ point[7] and two indirect bandgaps at about 1.2 eV and 0.8 eV at L and ~X points[8], whereas in the case of MgO, the energy gap of 7.8 eV is observed. ZnO naturally crystallizes in the wurtzite (hexagonal) structure, whereas CdO and MgO in the cubic rocksalt phase. Thus, in the case of ZnMgO and ZnCdO alloys, a phase separation issue is reported[9,10]. For (Zn,Cd)O, the researchers established pure wurtzite single crystalline up to 2% of Cd content[11]. The higher Cd concentration, up to 30 %, leads to mixture of wurtzite and zinc blende phases by low temperature cathodoluminescence mapping[9]. Whereas concentrations in the range of 32 % to 60 % coexistence of three crystal phases of wurtzite, zinc blende and rock salt. The full transition from the wurtzite to the rocksalt structure- takes place for above 62 % Cd content[11].

Random ZnCdO ternary alloys can be deposited using various techniques such as DC reactive magnetron sputtering[12], Molecular Beam Epitaxy (MBE)[13], sol-gel[14], metal oxide chemical vapor deposition (MOCVD)[9], and electrochemical deposition[13], etc. Digital alloys, or short-period superlattices (SLs), consisting of binary or ternary layers with a period of a few monolayers, have emerged as a solution for MBE growth of ternary or quaternary materials of various composition. In this way, it is possible to obtain a wide range of compositions without additional source cells and any laborious change of the cell temperature during growth. The number of papers about SLs - quasi-ternary alloys based on oxides is very limited. Good control of the energy gap in the case of quasi-ternary alloys short period superlattices[15–18] can give a chance to obtain semiconductor laser diodes or photodetectors for the dedicated spectral range. From an application point of view it is important to analyze influence of the high temperature on physical properties of ternary alloys. In this paper, we explore short period $\{CdO/ZnO\}_m$ superlattices, which were grown by plasma-assisted molecular beam epitaxy (PA-MBE) on *m*-oriented sapphire substrates, and we analyze optical properties of the as grown and annealed superlattices.

I. Experimental

The series of $\{ZnO/CdO\}_{30}$ superlattice samples were grown in a RIBER Compact 21 molecular beam epitaxy system equipped with oxygen plasma cells. MBE growths were performed on epi-ready *m*-plane (10-10) sapphire ($Al_2O_3$) substrates mounted on an indium-free molybdenum holder. Before the growth process, the sapphire substrates were chemically cleaned in an $H_2SO_4:H_2O_2$ (1:1) mixture. During the growth process, the radio-frequency (RF) power of the oxygen plasma was settled at about 400 W with an $O_2$ gas flow rate of 3 sccm. In the analyzed series of samples, the growth temperatures varied from 360°C to 550°C (Z360, Z400, Z450, Z500, and Z550 – numbers are the growth temperature of the sample).
The number of repeating ZnO and CdO pairs in SLs was 30 and the ZnO and CdO layers thickness was controlled by adjusting the deposition time of the ZnO and CdO sublayers. All samples also have a thin ZnO cap on top.

After the growth, rapid thermal processing (RTP) was performed (AccuThermo AW610 from Allwin21 Inc.) at 900 °C in pure oxygen environment ($O_2$).

Structural properties were investigated by high resolution X-ray diffraction (HR-XRD) with an X'Pert Pro-MRD Panalytical diffractometer using CuKα radiation.

The depth profiles of the Cd, Zn, and O elements in {CdO/ZnO}$_m$ SLs were investigated by the secondary ion mass spectrometry (SIMS) method using a CAMECA IMS6F system. A primary $Cs^+$ ion beam at an energy of 5.5 keV and a current of 50 nA was used. The $Cs^+$ ion beam covered the area of 150 μm × 150 μm and the secondary ions were collected from the central region with a diameter of 60 μm. The signals from the $^{114}CdCs^+$, $^{64}ZnCs^+$, and $^{16}OCs^+$ clusters were analyzed.

Optical properties of as deposited {ZnO/CdO}$_{30}$ SLs were analyzed in the wavelength range of 300 – 650 nm at room temperature (RT) using the two-channel Varian Cary 5000 spectrophotometer. For the photoluminescence (PL) experiments the samples were excited with the 330 nm line of an Argon laser in the temperature range of 10 – 250 K.

**2. Results**

2.1 Structural characterization

XRD measurements were performed to analyze changes in the crystallographic quality of the SLs structures due to annealing. Fig. 1 (a) and (b) present the high angle-resolution XRD spectra (2θ-ω) scans of SLs structures grown at different temperatures. Part (a) contains data on as grown and part (b) the structures after annealing. The main reflection peak (10.0) originating from the wurtzite structure is accompanied by satellite reflections from the superlattices in all as grown samples. The structures grown at lower temperatures have better crystallographic quality, as the satellite peaks up to sixth order are well visible, than the structures grown at higher temperatures. This fact is also confirmed by TEM analysis (published elsewhere)[19]. Interestingly, after annealing the satellite peaks, confirming the periodic crystal structure, are still observed only in the sample grown at 360°C. For the annealed samples grown in the temperatures of 400°C - 550°C single peak without satellites is observed, which suggests the partial or complete degradation of the periodic structure. This observation can be related to the interdiffusion of the Cd and Zn atoms at high temperatures. This conclusion is backed up with the results of Stachowicz et al.[20], which delivered data on interdiffussion coefficient in the highly disturbed ZnCdO random alloys annealed at relatively low temperature (650 $^o$C). The main 10.0 XRD peaks originating from the SLs are shifted for the annealed samples towards larger angles compared to the as grown samples. As it was mentioned above, the shift of the peaks may be associated with atom diffusion in the samples during annealing. The lattice parameters $a$ were calculated based on the HR XRD measurements and the results are listed in Table 1.

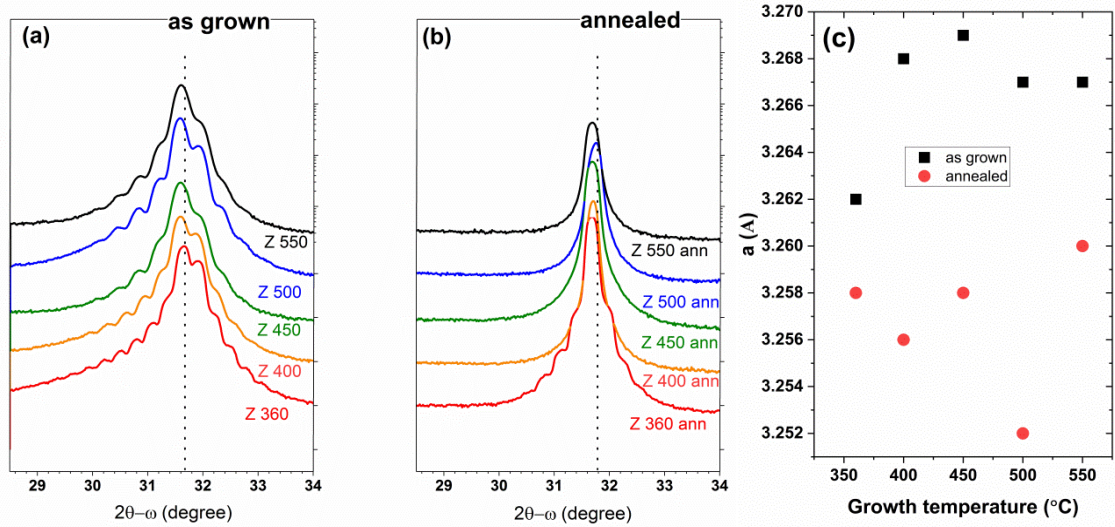

Fig. 1 High resolution X-ray diffraction 2θ/ω scans of {ZnO/CdO} superlattices (SL) on sapphire (a) as grown (b) after annealing at 900°C. (c) Average Lattice parameter for as grown SLs (black points) and for annealed samples (red points).

Table 1.

| Sample | $T_g$ [°C] | | $2\theta_{(100)}$ [°] | $a_{(100)}$ [Å] |
|---|---|---|---|---|
| Z360 | 360 | as grown | 31.64 | 3.262 |
| | | RTP 900 °C | 31.69 | 3.258 |
| Z400 | 400 | as grown | 31.59 | 3.268 |
| | | RTP 900 °C | 31.71 | 3.256 |
| Z450 | 450 | as grown | 31.58 | 3.269 |
| | | RTP 900 °C | 31.69 | 3.258 |
| Z500 | 500 | as grown | 31.60 | 3.267 |
| | | RTP 900 °C | 31.75 | 3.252 |
| Z550 | 550 | as grown | 31.60 | 3.267 |
| | | RTP 900 °C | 31.69 | 3.260 |

The Cd depth distribution in the as-deposited and annealed SLs was analyzed by SIMS (Fig. 2a). The Cd concentration was calculated from the ratio of CdCs+ and OCs+ signals as well as RSF (relative sensivity factor) calculated from the result of standard sample measurement. The periods of the as-deposited {ZnO/CdO} SLs grown at 360°C are clearly identifiable in SIMS depth profiles and correspond very well to the designed structure. On the surface, the Cd level is lowering because of the presence of a thin pure ZnO cap on the top of SLs. Fig. 2a shows the SIMS depth profiles for samples grown at various growth temperatures. It is almost impossible to detect individual CdO and ZnO layers in SLs due to limited depth resolution of SIMS measurement. However, the XRD data delivered a proof that SLs still exist. Perhaps the fact that Cd "concentration oscillation" are almost invisible in SIMS signals is due to an increase in the growth temperature which leads to an increase in Cd diffusion efficiency into the neighboring layers[21] or it can be correlated with some inhomogeneity of the samples.

The SIMS measurements were taken again after annealing at 900 °C in $O_2$ atmosphere for 5 min. The post-annealing SIMS depth profiles for all annealing samples look similar (Fig. 2 b) and the Cd concentration in the samples is almost homogenous indicating effective interdifussion of metallic sublattice elements into neighboring layers, which leads to thickness thinning of the individual layers making them indistinguishable for SIMS technique.

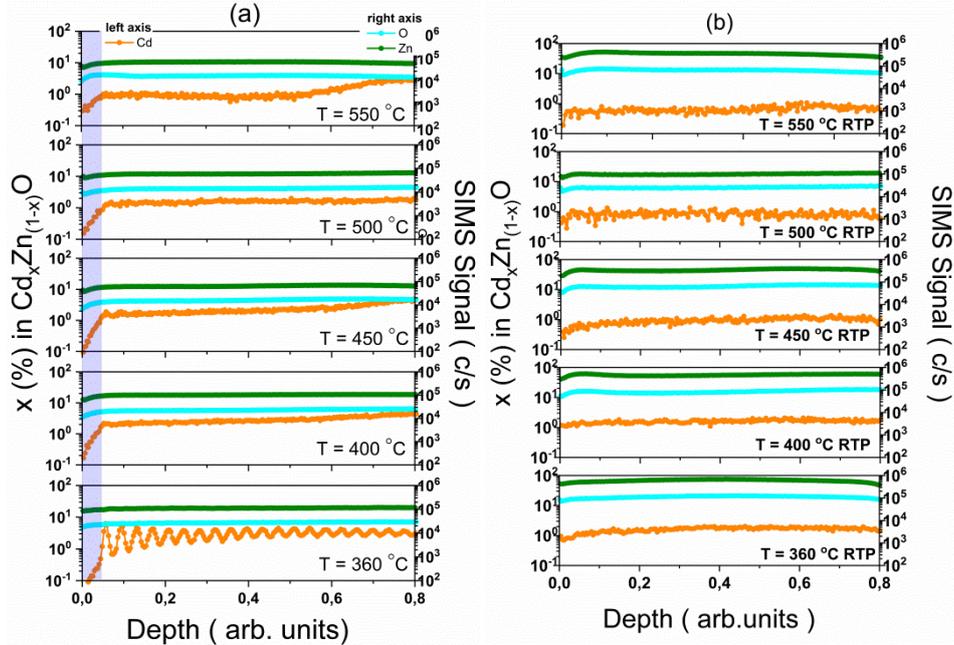

**Fig. 2.** SIMS depth profiles for structures prepared at various growth temperature (a) of as grown {ZnO/CdO}$_{30}$ SLs (b) annealed {ZnO/CdO}$_{30}$ SLs. Blue and green lines represents oxygen and zin signals-right axis; orange lines represent Cd content –left axis.

2.2 Optical properties

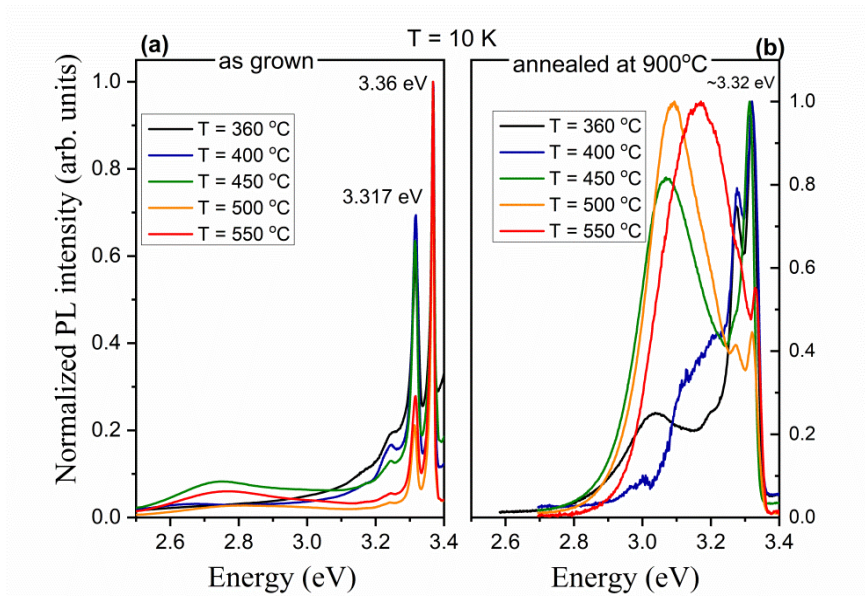

**Fig. 3** (a) Normalized photoluminescence spectra of as grown {ZnO/CdO}$_{30}$ SLs measured at low temperature (b) Normalized photoluminescence spectra of annealed {ZnO/CdO}$_{30}$ SLs

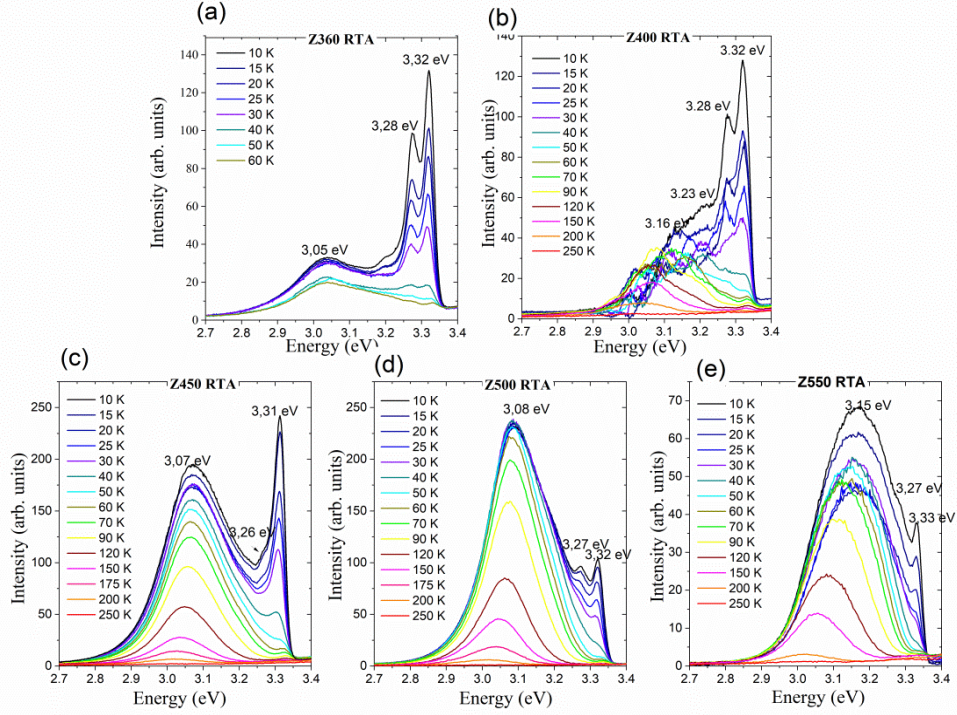

Fig. 3 presents the normalized PL spectra measured at low temperature (10 K) of the {ZnO/CdO}$_{30}$ SLs obtained at different growth temperatures. In all the as-grown SL samples (Fig. 3 (a)) the dominant emission is observed in the UV region at about 3.36 eV. The energy of the peak position suggests that it is probably related to the emission that comes from pure ZnO sublayers in SLs. The probability of radiative recombination in direct band gap materials (e.g.: ZnO) is higher than those in indirect band gap materials (e.g.: CdO). In temperature dependent PL (Fig 4(a-e)) when temperature increases the intensity of subjected emission peak strongly decreases, and a higher energy peak persist visible. Based on temperature dependent luminescence analysis, we identified the sharp PL peaks in the UV region (3.36-3.38 eV) originating from donor bound excitons D°X and free excitons (FX). As can be found in the literature, most of the ZnO films commonly exhibits the D°X emission around 3.36 eV.[22] The decrease in the intensity of the D°X peaks with the temperature and the simultaneous relative increase in the intensity of the FX peaks (observed in the Fig 5 (a)) can be understood in terms of thermal ionization of donors $D^0 \rightarrow D^+ + e$.

In all as-deposited samples the appearance of the peaks at the energy at about 3.317 eV is also detected. The origin of these peaks can be correlated with free electron to acceptor transition (FA), donor-acceptor pairs (DAP), and/or stacking faults. The intensity of the emission line centered at 3.317 eV at 10 K increases with temperature up to 30 K-40 K followed by a small decrease above 60 K. Such a behavior of the emission lines intensities up to 60 K can be also understood in terms of thermal ionization of donors $D^0 \rightarrow D^+ + e$, which increases the concentration of free electrons further promote – DAP intensity emission. Temperature dependent behavior of the peaks at about 3.317 eV (visible in all investigated samples) indicates that this can be correlated with DAP emission line (Fig. 5(b)).

A low intensity band in comparison to the UV region localized at about 2.6-3 eV is also visible in the investigated as-deposited samples (Fig. 3 (a)). The origin of this band can be correlated with some defects like oxygen vacancies ($V_o$) - usually detected in the green region at about ~2.5 eV. The deep emission at about 2eV is also identified as a zinc vacancies or oxygen interstitials[23]. In fact, however, we cannot clearly attribute the origin of this emission to any particular defect.

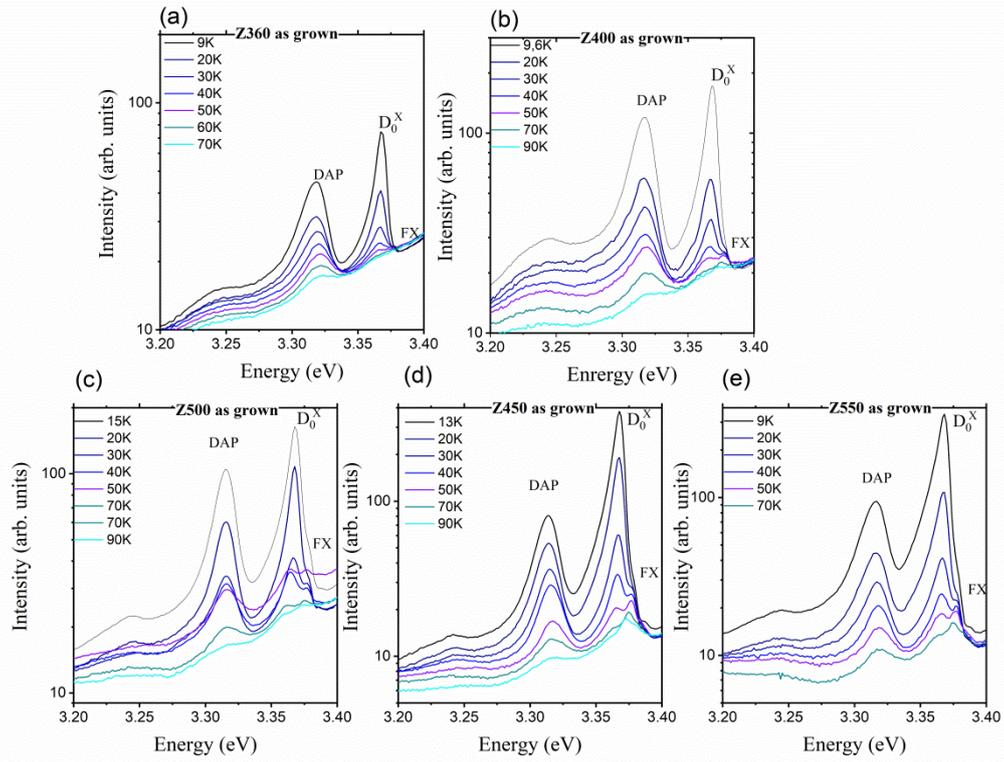

**Fig. 4.** Temperature-dependent PL spectra in NBE region in semi logarithmic scale of as grown {ZnO/CdO}$_{30}$ SLs obtained at various growth temperature (a) T = 360 $^{o}$C, (b) T = 400 $^{o}$C, (c) T = 450 $^{o}$C, (d) T= 500 $^{o}$C and (e) T= 550 $^{o}$C.

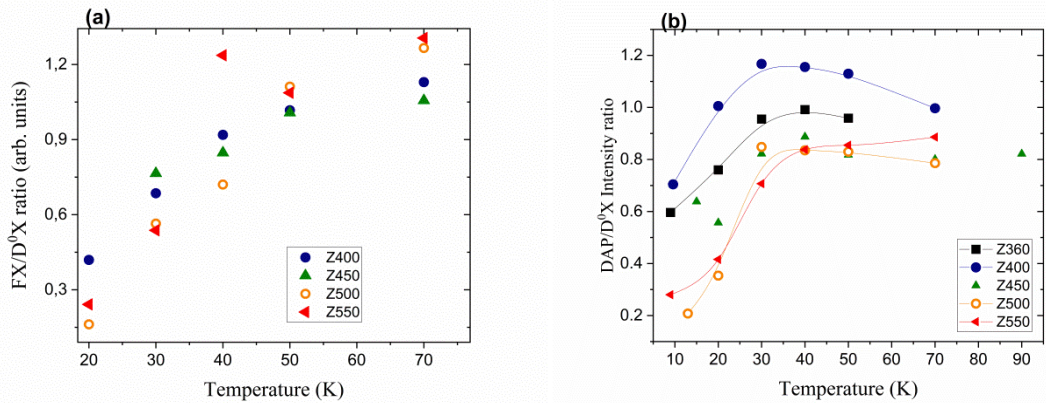

**Fig. 5** Intensity ratio of the free exciton peak FX to the donor bound exciton peak $D^0X$ measured at different temperatures. Different symbols correspond to the samples grown at different temperatures, respectively.

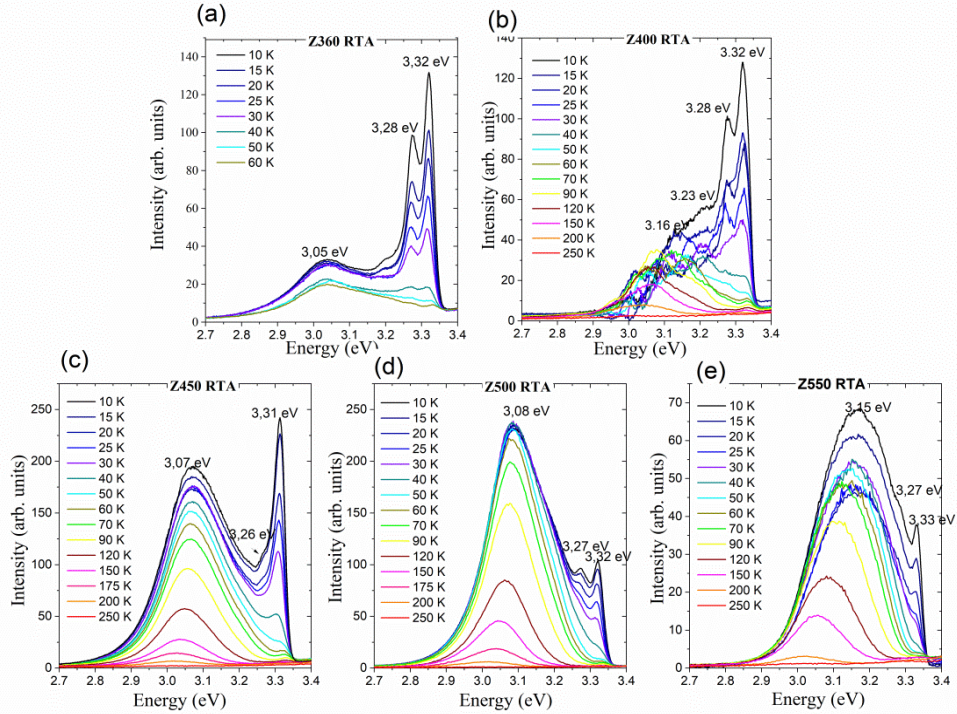

**Fig. 6.** Temperature-dependent PL spectra in semi logarithmic scale of annealed by RTA {ZnO/CdO}$_{30}$ SLs obtained at various growth temperature (a) T = 360 °C, (b) T = 400 °C, (c) T = 450 °C, (d) T= 500 °C and (e) T= 550 °C.

Figure 3 (b) shows the PL spectra of the {ZnO/CdO}$_{30}$ SLs annealed at 900 °C in an oxygen environment measured at low temperature (10 K) and normalized to the most intensive peak. In all annealed samples the near-band-edge (NBE) emission peaks at about 3.31-3.33 eV are observed and also a broad band of different intensity in relation to NBE is detected. In contrast to the as grown samples, the intensive PL band appears with a maximum located at about 3.1 eV. The luminescence spectra of the annealed structures strongly depend on the applied growth temperature. In particular, the relative intensity of the broad band to NBE main peaks changes significantly. In the case of samples grown at relatively high temperatures (samples Z500 and Z550), the intensity of the mentioned band is stronger than in the case of samples obtained at lower temperature regimes (samples Z360 and Z400) (see Fig 6 (a-e)).

For annealed ZnCdO film the NBE UV emission peaks are broadened and redshifted toward lower energy in comparison to as grown layers (Fig 7), which could be expected from the changes of the optical band gap derived from the absorption spectrum, which will be discussed later in this work. Due to annealing, the SLs structure is diffused out and as a consequence a ternary alloy is obtained. Shrinkage of the band-gap caused by Cd substitution suggests that Cd atoms are successfully incorporated into the ZnO structure, and incorporation of Cd can effectively change the band gap of ZnO, which causes a redshift. As the growth temperature increases, the PL peaks widen, and the intensity of defect related emission increases. We know from previously published TEM results[24] that the quality of the interfaces deteriorates with increasing growth temperature, thus increasing the number of structural defects, which may consequently positively affect the diffusion of Cd atoms during annealing. Thus, probably the change in the Cd diffusion mechanism influences the optical properties of the annealed samples[20].

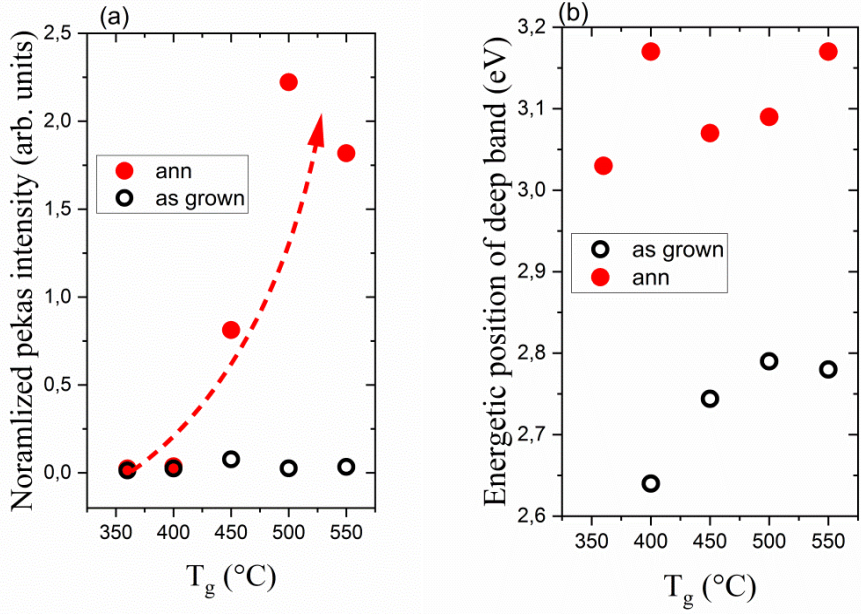

**Fig 7** (a) Normalized to $D^0X$ intensity at 10 K of deep emissions in as grown and annealed samples. (b) Energetic position of the deep emission for as grown and annealed samples.

Fig. 8 (a) shows the optical transmittance spectra of the as grown $\{ZnO/CdO\}_{30}$ SLs deposited at different growth temperature taken in the wavelength range of 300–650 nm.

The optical band gap ($E_g$) was estimated using the Tauc equation[25]:
$$\alpha h\nu = \beta(h\nu - E_g)^n, \quad (5)$$
where $h\nu$ is the photons energy, $E_g$ is the optical band gap energy, β is a constant related to the extent of the band tailing. It is a temperature independent constant that depends on the refractive index $n_o$ and correlated with $\beta$ by following equation[26]:

$$\beta = (4\pi c)\rho_0/n_0 E_U,$$

where $\rho_o$ is the electrical conductivity at absolute zero temperature, $E_U$ is the Urbach energy, and $n$ is a number characterizing the transition process, while $\alpha$ is the absorption coefficient. The absorption coefficient $\alpha$ can be calculated from the optical transmittance:
$$\alpha = -\frac{ln(T)}{d}, \quad (6)$$
where $d$ is the film thickness, $T$ is the transmittance.

For allowed direct transitions, $n = 1/2$, and for allowed indirect transitions, $n = 2$. The optical gaps $E_g$ could be determined from the extrapolation to zero of the linear regions of the $(\alpha h\nu)^2 = f(h\nu)$ (Fig. 8 (b)).

The band gap energy for as grown SLs samples decreases from 3.291 to 3.265 eV with increasing growth temperature. The growth temperature applied in MBE process affects the sticking coefficient of pure effused elements Cd, Zn and O. In connection with this, mutual amount of metals (Cd or Zn), and oxygen absorbed on the surface depend on temperature. This can influence the growth rate of CdO and ZnO with temperature, and hence the thickness of sublayers in SLs[24]. The thickness of sublayers in SLs affects the measured energy gap as it was previously observed in other superlattices systems[8,18,27].

Crystalline quality, grain size, carrier concentration and other factors may also slightly affect the change in the band gap of the samples[28]. A decrease in the direct band gap for pure CdO with an increase in the deposition temperature (from 300 to 412 °C) for films on glass and MgO(100) was observed by Metz et al.[29] In our case the crystal quality of the SLs improves with decreasing growth temperature and the value of the band gap shifts towards higher energies.

It was also shown that annealing of ZnCdO alloy influences the energy gap of these ternary alloy layers[30]. What is important in our study, is that the measured energy gaps of the annealed samples are lower than for as grown (Fig. 8) in all cases. As a result of the annealing process at high temperatures the initial SLs structures were diffused and consequently the CdZnO alloy layers were obtained. The consequence of the incorporation of Cd into the ZnO is the reduction of the measured energy gap[31],[32] and the red shift of the absorption edge in all annealed samples (Fig. 8). The applied growth temperature also influenced the energy gap of the annealed samples. The band-gap reduction of ZnO is mainly due to the formation of Cd 5s states below the conduction band (CB) edge as was mentioned by Sayed et. al.[33] It should be mentioned that in annealed samples prepared at higher growth temperatures, the high intensity defect luminescence band was observed.

To additionally check the crystal quality and also to have some information about the disorder, we calculated the Urbach energy. The Urbach energy changes in the range of 0,48 eV-0,26 eV for as grown samples and have a minimum at growth temperature of 500°C and varies from 0,41eV to 0,58eV for annealed samples (Fig.8 (c)). Obtained values of the Urbach energy for tested ternary alloys are in a good argument with previously reported data for CdZnO samples[34,35]. For every growth temperature, the measured Urbach energies are smaller for as grown samples in comparison to samples subjected to high temperatures in RTA process. Therefore, it indicates an increase of the disorder in the annealed samples.

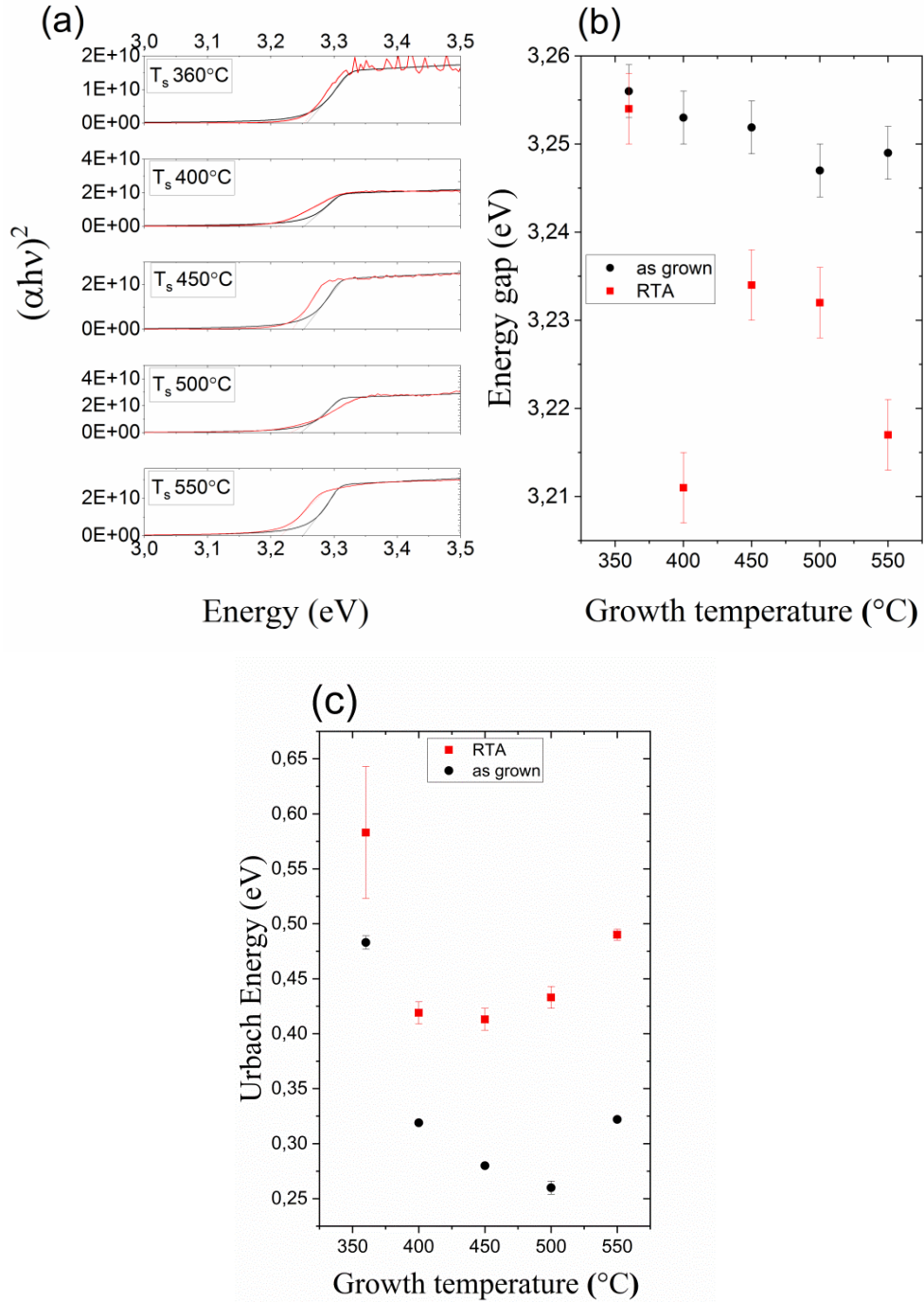

Fig. 8. (a) Transmission spectra of as-deposited (black lines) and annealed (red lines) {ZnO/CdO}$_{30}$ SLs grown at different temperature (360-550$^o$C); (b) Energy gaps values vs growth temperature (c) Urbach energy vs growth temperature .

## 3. Conclusions

Using the MBE method, we can make {CdO/ZnO} superlattices at different growth temperatures. The optical properties of the superlattice structures depend to a large extent on the temperature at which these structures were grown. Annealing of the superlattices at a high temperature of 900°C causes their partial or complete diffusion, and this process also depends on the growth temperature. Annealing also strongly affects the optical properties of these structures. In particular, the intensity and the spectral location of the defect related luminescence depend on the growth temperature. The changing position of the defect luminescence suggests the formation of another type of defects in these structures. The obtained results indicate that the growth temperature is one of the very important parameters that may affect future devices based on {CdO/ZnO} superlattices.

**Author statement**

**Ewa Przeździecka**: Conceptualization, Writing - Original Draft , Writing - Review & Editing, Project administration, Supervision **Anastasia Lysak**: Investigation, Methodology, Writing - Review & Editing. **Abinash Adhikari**: Investigation, Formal analysis. **Marcin Stachowicz**: Methodology, Investigation, Writing - Review & Editing **Rafał Jakieła**: Methodology, Investigation, Writing - Review & Editing **Khosravizadeh Zeinab**: Investigation **Piotr Sybiklski**: Investigation, **Adrian Kozanecki:** Writing - Review & Editing

**Funding** This work was supported in part by the Polish National Science Center , Grants No . 2021/41/B/ST5/00216 .

**Declaration of competing interest**

The authors declare that they have no known competing financial interests or personal relationships that could have appeared to influence the work reported in this paper. E. Przezdziecka reports financial support was provided by National Science center Poland. A. Lysak reports financial support was provided by National Science center Poland. A. Adhikari reports financial support was provided by National Science center Poland. M. Stachowicz reports financial support was provided by National Science center Poland.  A. Wierzbicka reports financial support was provided by National Science center Poland. A. Kozanecki reports financial support was provided by National Science center Poland.

**Data availability**

Data will be made available on request.